\documentclass[
aip,
amsmath,amssymb,
preprint,%
]{revtex4-1}

\usepackage{graphicx}
\usepackage{dcolumn}
\usepackage{bm}
\usepackage[mathlines]{lineno}

\usepackage[utf8]{inputenc}
\usepackage[T1]{fontenc}
\usepackage{mathptmx}
\usepackage{etoolbox}

\makeatletter
\def\@email#1#2{%
 \endgroup
 \patchcmd{\titleblock@produce}
  {\frontmatter@RRAPformat}
  {\frontmatter@RRAPformat{\produce@RRAP{*#1\href{mailto:#2}{#2}}}\frontmatter@RRAPformat}
  {}{}
}%
\makeatother
\begin{document}
\preprint{}

\title[]{On the numerical simulations of electric field-enhanced solid-state electro-aerodynamic thrusters enabled with an insulated electrode\footnote{Work in progress}}
\author{Hisaichi SHIBATA}
 \affiliation{Graduate School of Science and Technology, Keio University.}
 \email{altair@keio.jp (H.S.).}
\author{Takahiro NOZAKI}%
\affiliation{Department of System Design Engineering, Keio University. 
}%

\date{\today}

\begin{abstract}
A solid-state electro-aerodynamic propulsion system applies a high electric potential difference between two electrodes to ionize air within the resulting electric field, accelerating the ions to generate thrust. Previously, the so-called ``decoupled thrusters'' have been proposed, in which ion generation and acceleration are spatially decoupled. Here, we argue that such decoupling is infeasible from the perspective of electric field lines if the diffusion effects are ignored, and consequently the driving method (e.g., dielectric barrier discharge) for the additional electrodes can be improved. Specifically, the use of auxiliary electrodes that do not discharge but induce ionization and attachment can be appropriate. This concept was named the electric field-enhanced solid-state electro-aerodynamic propulsion system. Furthermore, through numerical simulations, this concept showed improvement in thrust density while maintaining the thrust-to-power ratio.
\end{abstract}

\maketitle

Solid-state electro-aerodynamic propulsion systems (SSEPs) \cite{xu2018flight, gomez2023model, gomez2024order, masuyama2013performance, gilmore2015electrohydrodynamic, xu2019higher, xu2019dielectric, gomez2021performance} utilize ambient air to obtain net thrust and can realize nearly silent drones.
However, SSEPs have a low thrust density \cite{masuyama2013performance, gilmore2015electrohydrodynamic}.
In order to ameliorate this, decoupled thrusters have been proposed \cite{xu2019dielectric, gomez2021performance}.
In their work, it was argued that they aimed to decouple ion generation from ion acceleration.
Specifically, they adopted dielectric barrier discharge (DBD) driven by an alternative current of up to 100 kHz to ionize ambient air prior to acceleration by a direct-current electric field.

On the other hand, the authors of the present study have enabled the numerical modeling of atmospheric pressure non-equilibrium low-temperature plasma that appears in SSEPs \cite{shibata2025integral, shibata2016performance, shibata2022novel, shibata2025anemone}.
We adopt the method of characteristics, which can trace the physical quantities, e.g., negative and positive ion densities, along the electric field lines \cite{shibata2025integral}.
During this formulation, we have investigated that the decoupled thrusters do not decouple the ionization and acceleration procedures if we focus on an electric field line.
If a completely decoupled line exists, the line can ionize the ambient air molecules but cannot accelerate it or vice versa; hence, it cannot produce net thrust.
Moreover, we consider that the DBD discharge current perpendicular to the thrust axis does not contribute to the net thrust; hence, it is a kind of reactive power, resulting in the lower thrust-to-power ratio.

On the basis of the above new insights, we propose a new class of SSEPs, which we named electric field-enhanced SSEPs (EFE-SSEPs).
The physical structure of EFE-SSEPs is the same as in previous studies, but the high-frequency power supply has been replaced with a direct-current (DC) power supply.
This new concept ensures that the discharge between the emitter and the third electrode, which was previously a DBD electrode, does not occur but can enhance the electric field around the emitter; hence, this can contribute to efficient and powerful ionization.

We assume an impulsive moment just after the discharge begins.
This justifies ignoring the surface charge effects on the third electrode.
The steady-state three-component (i.e. electrons, positive and negative ions) plasma fluid equations are given by
\begin{eqnarray}
    - \nabla \cdot \bm{J}_e &=&  \left( \alpha - \eta \right) |\bm{J}_e|, \\
\label{eqn:1dl}
     \nabla \cdot \bm{J}_+ &=&  \alpha |\bm{J}_e|, \\
\label{eqn:2dl}
    - \nabla \cdot \bm{J}_- &=&  \eta |\bm{J}_e|, 
\label{eqn:3dl}
\end{eqnarray}
where the current density fluxes are given by 
\begin{eqnarray}
\bm{J}_e &=& \mu_e \bm{E} \rho_e, \\
\bm{J}_+ &=& \mu_+ \bm{E} \rho_+, \\
\bm{J}_- &=& \mu_- \bm{E} \rho_-,
\end{eqnarray}
where $\rho_e$, $\rho_+$ and $\rho_-$ are the charge density of electrons, positive and negative ions, respectively. $\mu_e > 0$ and $\mu_\pm > 0$ are the mobility of electrons, positive and negative ions, and these are assumed to be constant, respectively. $\alpha$ and $\eta$ are the electron impact ionization coefficient and the electron attachment coefficient as functions of the electric field strength $E = |\bm{E}|$ and the number density of the air (local reduced field).

These drift-reaction equations are solved with the Poisson equation,
\begin{equation}
    \nabla \left( \varepsilon_r \nabla \phi \right) = -\frac{1}{\epsilon_0} \left( \rho_+ - \rho_- - \rho_e \right),
\end{equation}
and 
\begin{equation}
    \bm{E} = -\nabla \phi,
\end{equation}
where $\varepsilon_0 (=8.8\times 10^{-12} \ \mathrm{F/m})$ is the permittivity of the air, and $\varepsilon_r$ is the relative permittivity.

The boundary conditions for the continuity equations are if $\bm{E} \cdot \hat{\bm{n}} \ge 0$ ($\hat{\bm{n}}$ is the unit outward normal vector on the boundaries), 
\begin{eqnarray}
    \rho_+ &=& 0, 
\end{eqnarray}
and otherwise we have
\begin{eqnarray}
    \mu_e \rho_e &=& \gamma \mu_+  \rho_+, \\
    \rho_- &=& 0,
\end{eqnarray}
where $\gamma (=0.128)$ is the secondary electron emission coefficient.
   
The boundary conditions for the Poisson equation are, on the far-field boundaries,
\begin{eqnarray}
    \hat{\bm{n}} \cdot \nabla \phi &=& 0,
\end{eqnarray}
on the emitter, 
\begin{eqnarray}
    \phi &=& 0,
\end{eqnarray}
on the collector,
\begin{eqnarray}
    \phi &=& V_c,
\end{eqnarray}
and on the third electrode,
\begin{eqnarray}
    \phi &=& V_3,
\end{eqnarray}
where we assume $V_c > 0$ and $V_3 \ge 0$ in this study.

The total power consumption $P$ is calculated by
\begin{eqnarray}
    P = I_c V_c + I_3 V_3, 
\end{eqnarray}
where $I_c$ and $I_3$ are the current on the collector and the third electrode, respectively, which are calculated by
\begin{eqnarray}
    I = \oint_{\Gamma} \left( \bm{J}_e + \bm{J}_+ + \bm{J}_- \right) \cdot \hat{\bm{n}} \mathrm{d}\Gamma,
\end{eqnarray}
where $\Gamma$ represents the electrode boundary.

Finally, we approximate the thrust density by the Coulomb force given by
\begin{eqnarray}
    T \approx \iint_\Omega \left( \rho_+ - \rho_- \right) \bm{E} \mathrm{d}\Omega,
\end{eqnarray}
where $\Omega$ represents the computational domain.

To solve the above coupled equations for charge densities, we adopt the formulation in a previous work by Shibata and Nozaki \cite{shibata2025integral}.
Specifically, we solve those with the finite element method for the Poisson equation and with the method of characteristics for the current continuity equations.

\begin{figure}
    \centering
    \includegraphics[width=0.99\linewidth]{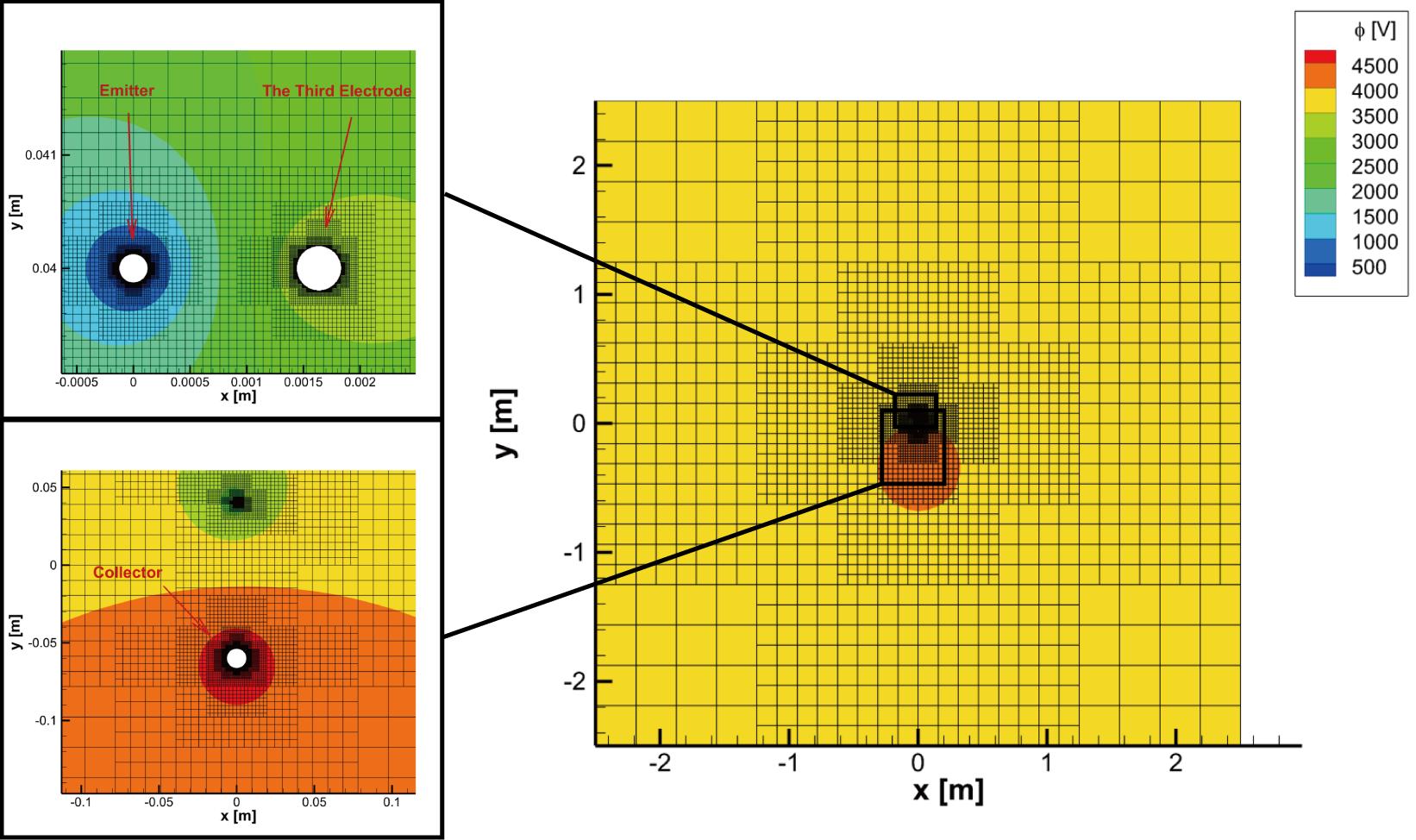}
    \caption{Computational grid and the Laplace field on EFE-SSEP's geometry (electrode arrangement). This is when $V_3 = 3.5$ kV and $V_c = 5.0$ kV.}
    \label{fig:geometry}
\end{figure}
We adopted the same electrode arrangement of the SSEP from a previous work \cite{xu2019dielectric} (Figure \ref{fig:geometry}), but assume that the dielectric on the third electrode is infinitesimally thin and has a relative permittivity $\epsilon_r = 1$, and the gap between the third and emitter electrodes was set at approximately 1.3 mm. We fixed the applied voltage of the third electrode to 0 or 3,500 volts and grounded the applied voltage of the emitter electrode to 0 volts, while we increased the applied voltage of the collector electrode from 5000 volts to 60,000 volts with the step of 1,000 volts. 
We traced electric field lines from the emitter electrode. We eliminated extremely long lines during the procedure. We assume atmospheric pressure air \cite{ferreira2019simulation}.

\begin{figure}
    \centering
    \includegraphics[width=\linewidth]{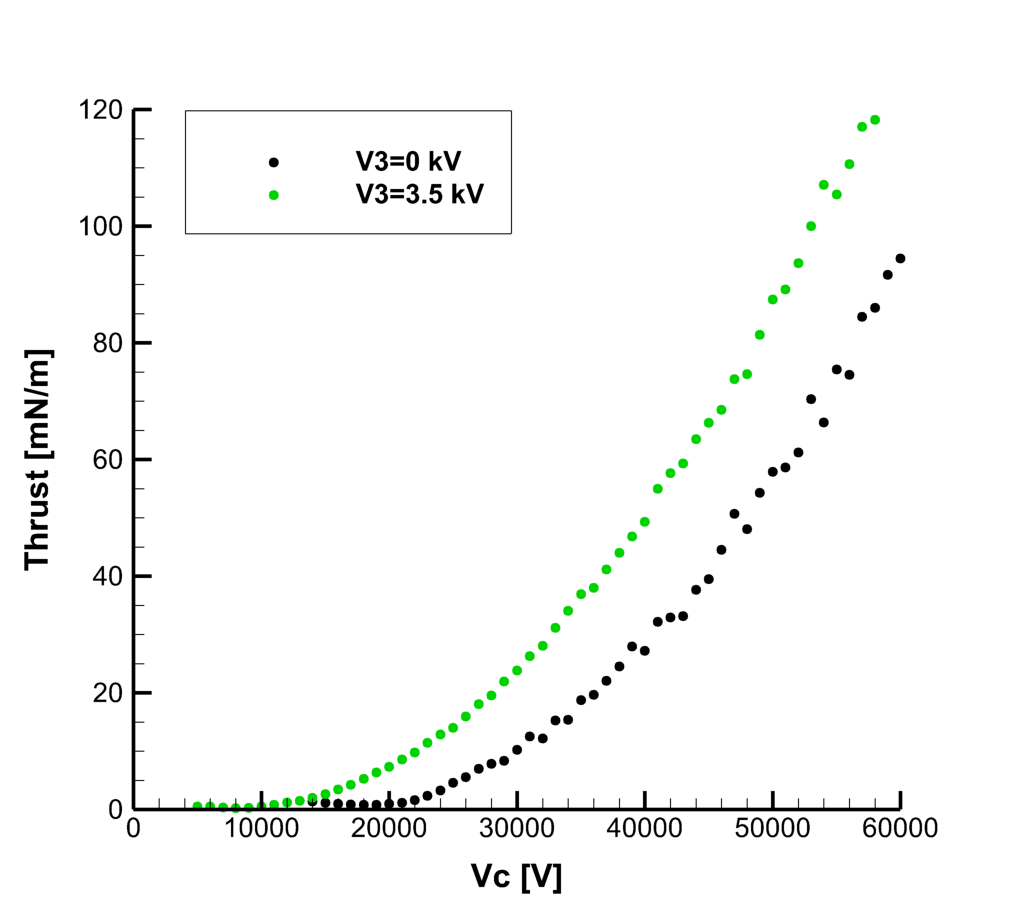}
    \caption{Thrust-voltage characteristics of the proposed SSEP.}
    \label{fig:tvc}
\end{figure}

\begin{figure}
    \centering
    \includegraphics[width=\linewidth]{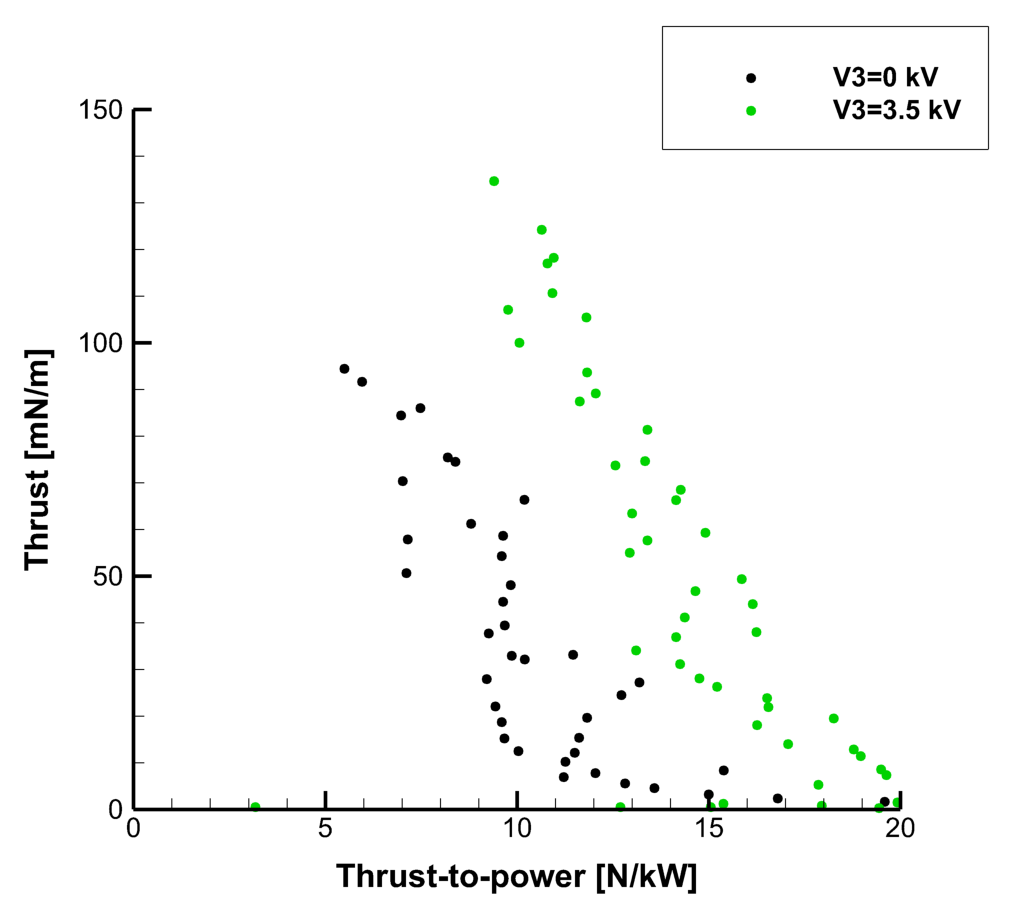}
    \caption{Thrust vs. thrust-to-power characteristics of the proposed SSEP.}
    \label{fig:t_and_tbyp}
\end{figure}

\begin{figure}
    \centering
    \includegraphics[width=\linewidth]{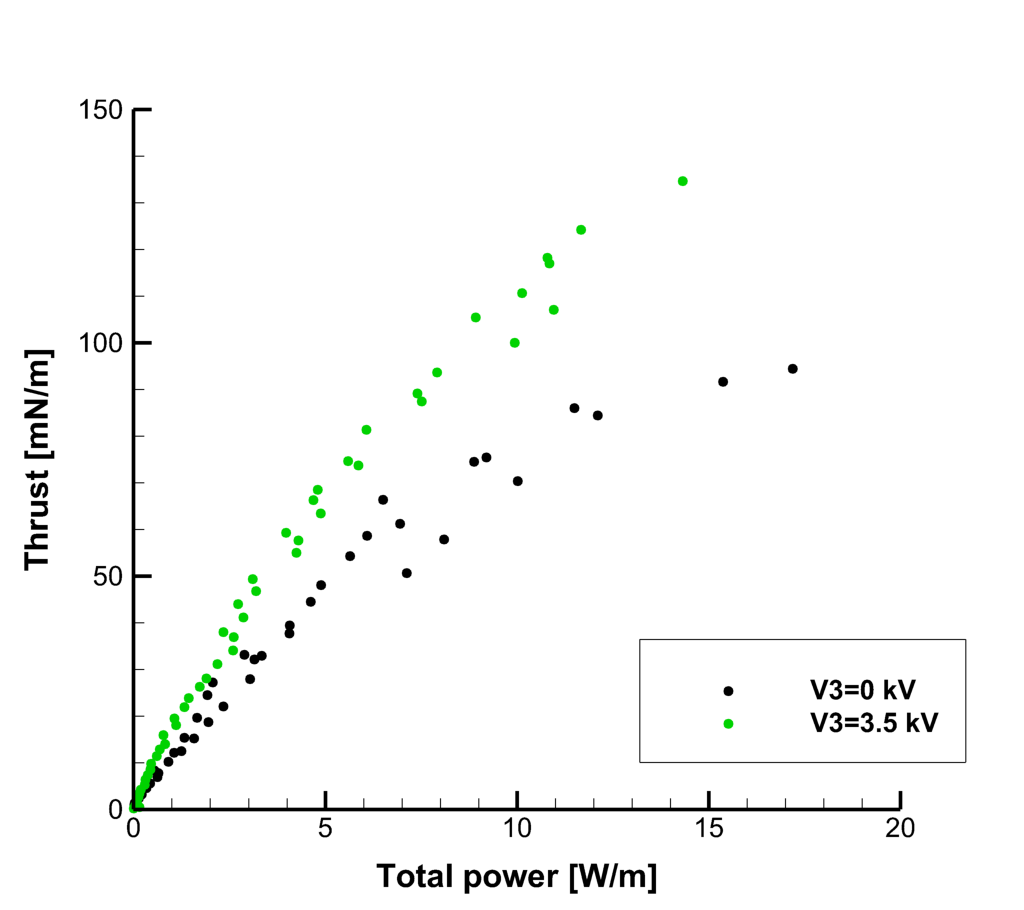}
    \caption{Thrust vs. power consumption of the proposed SSEP.}
    \label{fig:tpc}
\end{figure}

\begin{figure}
    \centering
    \includegraphics[width=0.99\linewidth]{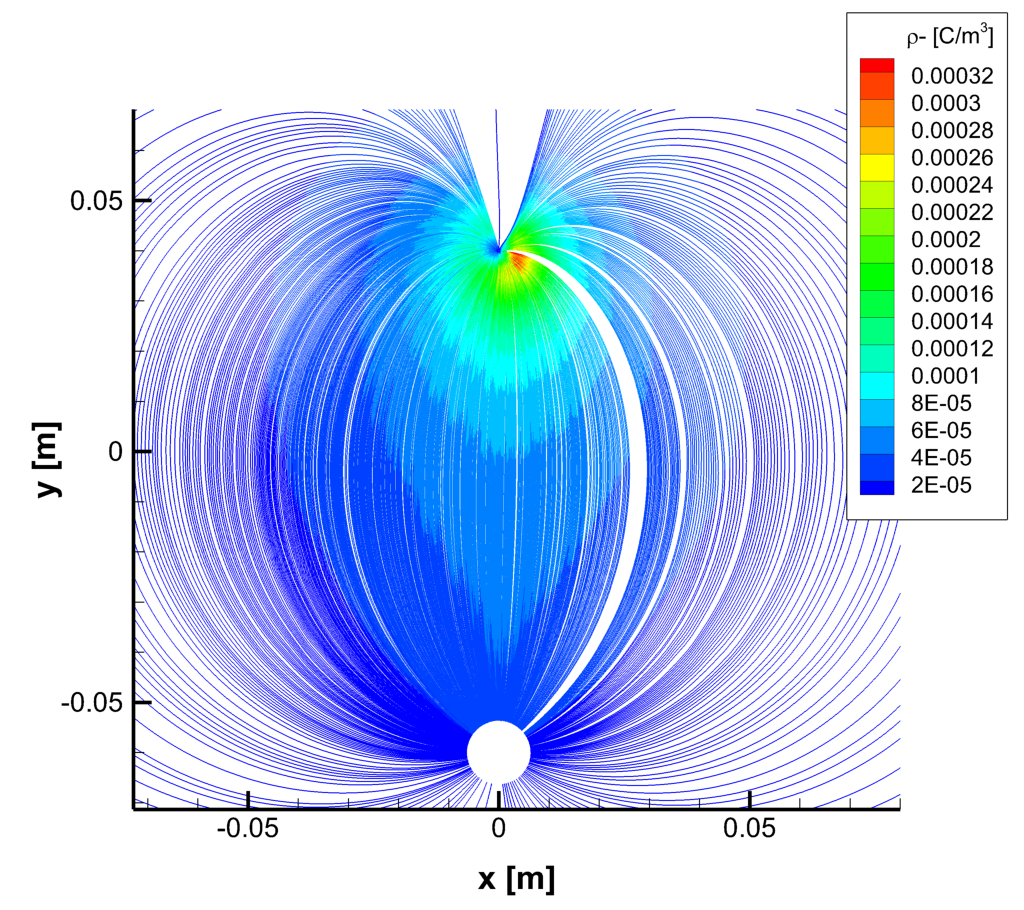}
    \caption{Negative ion density distribution on the electric field lines ($V_c = 60$ kV and $V_3 = 3.5$ kV).}
    \label{fig:neg_ion_1}
\end{figure}

\begin{figure}
    \centering
    \includegraphics[width=0.99\linewidth]{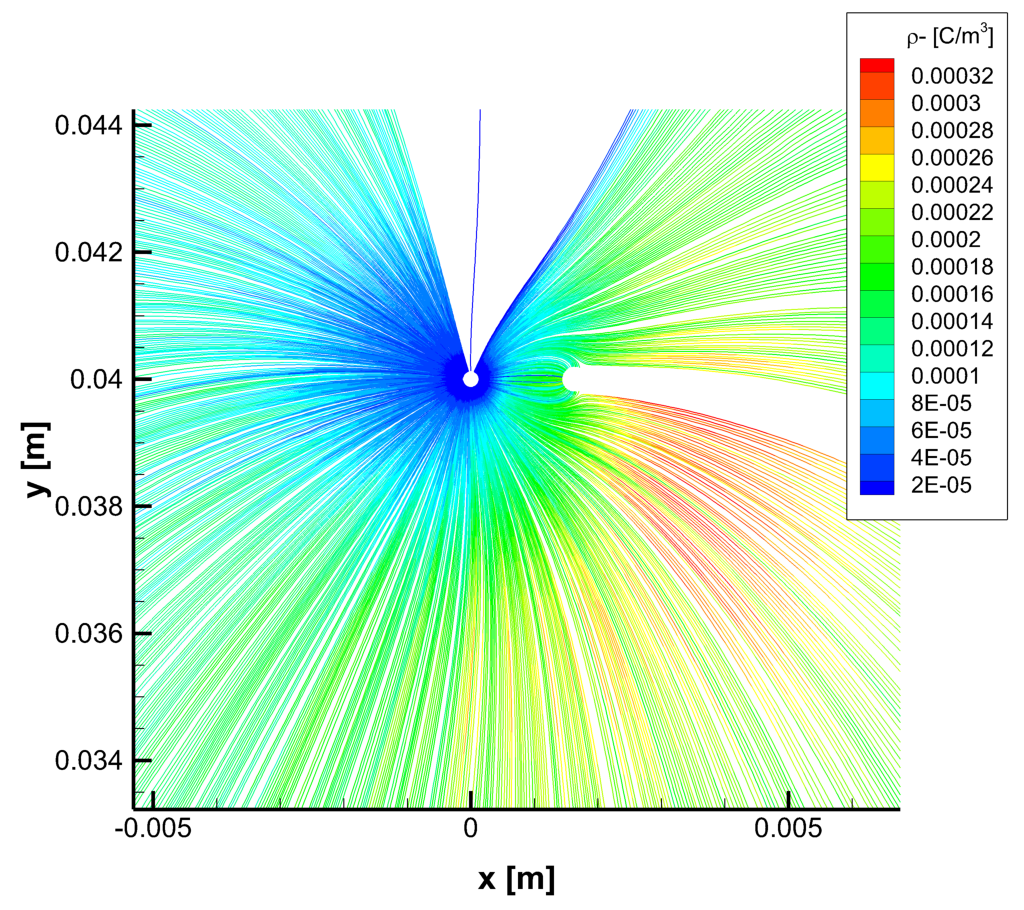}
    \caption{Negative ion density distribution on the electric field lines ($V_c = 60$ kV and $V_3 = 3.5$ kV, enlarged).}
    \label{fig:neg_ion_2}
\end{figure}

Figure \ref{fig:tvc} shows the thrust of the SSEP as a function of the applied voltage $V_c$.
Although we handled the negative corona, the characteristics show good agreement with the previous study (see the DBD off case in Figure 2 of \cite{xu2019dielectric}).
Figure \ref{fig:t_and_tbyp} shows the thrust as a function of the thrust to power ratio. The thruster performance is remarkably high when $V_3$ experiences a non-zero applied voltage.
Figure \ref{fig:tpc} shows the thrust of the SSEP as a function of total power consumption $P$. Even from the viewpoint of the total power, the proposed driven method can improve thruster performance.
Figures \ref{fig:neg_ion_1} and \ref{fig:neg_ion_2} show a negative ion density distribution on electric field lines.
The density is relatively high around the third electrode; hence, we can confirm that the third electrode contributes to the enhancement of the ionization and attachment.

We have investigated the characteristics of EFE-SSEPs, a new class of SSEPs, which contain an additional third electrode to enhance the electric field around the emitter electrode, with the aim of contributing to efficient and powerful ionization and consequently attachment.
Our numerical simulation revealed that it can significantly improve the thrust density while maintaining the total thrust-to-power ratio, unlike the previously proposed decoupled thrusters \cite{xu2019dielectric}.
Specifically, this newly proposed concept simultaneously achieved the thrust density of 50 mN/m and the thrust-to-power ratio of 15 N/kW, which marked the state-of-the-art performance.
In future work, we plan to experimentally validate this performance, including transient phenomena caused by surface charge deposition.

\begin{acknowledgements}
This work is based on results obtained from a project,
JPNP14004, commissioned by the New Energy and Industrial Technology Development Organization (NEDO).
\end{acknowledgements}

\section*{Data Availability}
The data that support the findings of this study are available from the corresponding author upon reasonable request.

\bibliography{aipsamp}

\end{document}